\documentclass{article}
\usepackage{amssymb}
\usepackage{amsmath}

\input{tcilatex}
\begin{document}

\begin{center}
\textbf{A simple approach to identify systems of nonlinear recursions
featuring solutions whose evolution is explicitly ascertainable and which
may be asymptotically isochronous as functions of the independent variable
(a ticking time)}

\bigskip

\textbf{Francesco Calogero}

Physics Department, University of Rome "La Sapienza", Rome, Italy

Istituto Nazionale di Fisica Nucleare, Sezione di Roma, Rome, Italy

Istituto Nazionale di Alta Matematica, Gruppo Nazionale di Fisica
Matematica, Italy

francesco.calogero@uniroma1.it, francesco.calogero@roma1.infn.it

\bigskip
\end{center}

\textbf{Abstract}

In this paper we describe a procedure which allows to identify new systems
of nonlinear recursions featuring solutions whose evolution is \textit{%
explicitly }ascertainable and which may be \textit{asymptotically isochronous%
} as functions of the independent variable (considered as a \textit{ticking
time}).

\bigskip

\textbf{Introduction}

This paper is a follow-up to $3$ previous papers [1,2,3 ] put by myself on 
\textbf{arXiv} recently, and might well be my last scientific publication
(see the last sentence in [3]).

In the \textit{second} and \textit{third} of those $3$ papers I used the 
\textit{quite standard} approach which starts from \textit{linear} ODEs or
recursions, themselves of course \textit{explicitly} \textit{solvable} (here
and below I confine my attention to \textit{autonomous} differential
equations or recursions); and then identifies, via a \textit{nonlinear}
change of dependent variables, \textit{new nonlinear} differential equations
or recursions, which inherit some of the solvability features of the \textit{%
linear} systems.

But in the \textit{first} of those $3$ papers I used instead a different
approach to investigate some \textit{simple nonlinear} recursions. I started
from the very simple single \textit{nonlinear }recursion $y\left( n+1\right)
=\left[ y\left( n\right) -1\right] ^{2}$ (hereafter $n=0,1,2,3,...$ is the
independent variable: think of it as a \textit{ticking time}), which, while 
\textit{not} being (to the best of my knowledge) \textit{explicitly}
solvable---for an arbitrary initial datum $y\left( 0\right) $---in terms of
known functions, is in fact so simple to allow a rather detailed description
of its evolution. For instance, for (almost: see [1]) \textit{all} initial
data in the interval $0<y\left( 0\right) <1$, this evolution entails that at
every step the dependent variable $y\left( n\right) $ gets closer,
successively, to the $2$ extremes of that interval, approaching steadily one
of its $2$ trivial solutions---periodic with period $2$---$y\left( n\right) =%
\left[ 1\pm \left( -1\right) ^{n}\right] /2$. In that paper I have then
shown how, starting from this elementary finding, one can identify other 
\textit{nonlinear} recurrences which, while also \textit{not} being
themselves \textit{explicitly} solvable in terms of known functions, do
allow rather detailed predictions about the behavior of their solutions: for
instance, they all \textit{feature} the property of \textit{asymptotic
isochrony} with period $2$, namely all solutions starting from a certain
identifiable range of initial data $y\left( 0\right) $ feature the property $%
y\left( n+2\right) -y\left( n\right) \rightarrow 0$ as $n\rightarrow \infty $%
.

In the present paper I indicate how this approach can be similarly used to
identify possibly interesting \textit{systems} of \textit{nonlinearly-coupled%
} recursions, such that the rather detailed behavior of their solution can
be predicted even though their \textit{explicit} solutions are presumably
unattainable. For simplicity I confine my treatment to the transition from
the \textit{single} (\textit{nonlinear} but \textit{manageable}) recursion
discussed above, to a system of \textit{only} $2$ nonlinearly-coupled
recursions, which shall then turn out to be as well \textit{manageable};
various possibility to extend/generalize this approach are of course obvious.

\bigskip

\textbf{The transition from a single recursion to a system of }$2$\textbf{\
nonlinearly-coupled recursions}

Our starting point are just $2$ versions of the simple nonlinear recursion
discussed above:%
\begin{equation}
y_{j}\left( n+1\right) =\left[ y_{j}\left( n\right) -1\right]
^{2}~,~~~j=1,2~.  \label{1}
\end{equation}

\textbf{Notation}: as already mentioned, the independent variable $n$ takes 
\textit{all nonnegative integer} values, $n=0,1,2,3,...$; the indices $j$
and (see below) $k$ take the $2$ values $1$ and $2$ (and, see below, the
index $\ell $ shall take the $3$ values $\ell =1,2,3$); we shall confine
attention to the case when the $2$ dependent variables $y_{j}\left( n\right) 
$ (and, likewise, $x_{j}\left( n\right) $: see below) are \textit{real
numbers}, as well as all the \textit{coefficients} introduced
below---although an interesting property of \textit{recursions} of the kind
we consider is that consideration might be restricted to \textit{rational}
numbers (for the marginal, but nevertheless interesting, difference among
the $2$ cases of a \textit{real} numbers or \textit{rational} numbers
context, see [1]; of course extensions to more general contexts---for
instance, from\textit{\ numbers} to\textit{\ noncommuting} quantities such
as \textit{matrices}---shall eventually also be of future interest). $%
\blacksquare $

So, we now introduce the following \textit{change of variables}, from the $2$
dependent variables $y_{k}\left( n\right) $ to $2$ new dependent variables $%
x_{j}\left( n\right) $:%
\begin{equation}
x_{j}\left( n\right) =a_{j}+b_{j1}y_{1}\left( n\right) +b_{j2}y_{2}\left(
n\right) ~;  \label{2}
\end{equation}%
we have thus introduced $2+2\cdot 2=6$ \textit{a priori arbitrary}
parameters $a_{j}$ and $b_{jk}$. Then, inverting eq. (2), we get%
\begin{equation}
y_{j}\left( n\right) =\alpha _{j}+\beta _{j1}x_{1}\left( n\right) +\beta
_{j2}x_{2}\left( n\right) ~.  \label{3}
\end{equation}%
The $6$ parameters $\alpha _{j}$ and $\beta _{jk}$ are of course simply
expressed in terms of the $6$ parameters $a_{j}$ and $b_{jk}$: \ 
\begin{eqnarray}
\alpha _{1} &=&\left( -a_{1}b_{22}+a_{2}b_{11}\right) /B~,  \notag \\
\alpha _{2} &=&\left( a_{1}b_{21}-a_{2}b_{11}\right) /B~,  \notag \\
\beta _{11} &=&b_{22}/B~,  \notag \\
\beta _{12} &=&-b_{12}/B~,  \notag \\
\beta _{21} &=&-b_{21}/B~,  \notag \\
\beta _{22} &=&b_{11}/B~,  \notag \\
B &=&b_{11}b_{22}-b_{12}b_{21}~.  \label{4}
\end{eqnarray}

Hence, from eqs. (1) and (3), we get 
\begin{subequations}
\label{5}
\begin{eqnarray}
&&\beta _{j1}x_{1}\left( n+1\right) +\beta _{j2}x_{2}\left( n+1\right) = 
\left[ \alpha _{j}+\beta _{j1}x_{1}\left( n\right) +\beta _{j2}x_{2}\left(
n\right) -1\right] ^{2}-\alpha _{j}=  \notag \\
&&\left( \alpha _{j}\right) ^{2}-\alpha _{j}+2\alpha _{j}\left[ \beta
_{j1}x_{1}\left( n\right) +\beta _{j2}x_{2}\left( n\right) -1\right] +\left[
\beta _{j1}x_{1}\left( n\right) +\beta _{j2}x_{2}\left( n\right) -1\right]
^{2}=  \notag \\
&&A_{j}+B_{j1}x_{1}\left( n\right) +B_{j2}x_{2}\left( n\right) +C_{j1}\left[
x_{1}\left( n\right) \right] ^{2}+C_{j2}\left[ x_{2}\left( n\right) \right]
^{2}+C_{j3}x_{1}\left( n\right) x_{2}\left( n\right) ~,  \notag \\
&&  \label{5a}
\end{eqnarray}%
where 
\begin{eqnarray}
&&A_{j}=\left( \alpha _{j}\right) ^{2}-3\alpha _{j}+1~,  \notag \\
&&B_{jk}=2\left( \alpha _{j}-1\right) \beta _{jk}~,  \notag \\
&&C_{jk}=\left( \beta _{jk}\right) ^{2}~,~~~C_{j3}=2\beta _{j1}\beta _{j2}~;
\label{5b}
\end{eqnarray}%
we have thus obtained the system of $2$ recurrences 
\end{subequations}
\begin{eqnarray}
&&\beta _{j1}x_{1}\left( n+1\right) +\beta _{j2}x_{2}\left( n+1\right)
=A_{j}+B_{j1}x_{1}\left( n\right) +B_{j2}x_{2}\left( n\right)  \notag \\
&&+C_{j1}\left[ x_{1}\left( n\right) \right] ^{2}+C_{j2}\left[ x_{2}\left(
n\right) \right] ^{2}+C_{j3}x_{1}\left( n\right) x_{2}\left( n\right) ~.
\label{6}
\end{eqnarray}

And finally this system can clearly be reformulated as follows:%
\begin{eqnarray}
&&x_{j}\left( n+1\right) =D_{j}+E_{j1}x_{1}\left( n\right)
+E_{j2}x_{2}\left( n\right) +  \notag \\
&&F_{j1}\left[ x_{1}\left( n\right) \right] ^{2}+F_{j2}\left[ x_{2}\left(
n\right) \right] ^{2}+F_{j3}x_{1}\left( n\right) x_{2}\left( n\right) ~,
\label{7}
\end{eqnarray}%
with%
\begin{eqnarray}
D_{1} &=&A_{1}\beta _{22}-A_{2}\beta _{21}~,~~~D_{2}=-A_{1}\beta
_{21}+A_{2}\beta _{11}~,  \notag \\
E_{11} &=&B_{11}\beta _{22}-B_{21}\beta _{12}~,~~~E_{12}=B_{12}\beta
_{21}-B_{22}\beta _{12}~,  \notag \\
E_{21} &=&-B_{11}\beta _{21}+B_{21}\beta _{11}~,~~~E_{22}=-B_{12}\beta
_{21}+B_{22}\beta _{11}~,  \notag \\
F_{11} &=&C_{11}\beta _{22}-C_{21}\beta _{12}~,~~~F_{12}=C_{12}\beta
_{22}-C_{22}\beta _{12}~,  \notag \\
F_{21} &=&-C_{12}\beta _{21}+C_{21}\beta _{11}~,~~~F_{22}=-C_{12}\beta
_{21}+C_{22}\beta _{11}~,  \notag \\
F_{13} &=&-C_{13}\beta _{22}+C_{23}\beta _{12}~,~~~F_{23}=-C_{13}\beta
_{21}+C_{23}\beta _{11}~.  \label{8}
\end{eqnarray}

This system of $2$ recurrences for the $2$ dependent variables $x_{j}\left(
n\right) $ features $2+2\cdot 2+2\cdot 3=12$ coefficients $D_{k}$, $E_{jk}$
and $F_{j\ell }$, which are \textit{explicitly} expressed via the formulas
(8), (5b) and (4) in terms of the $6$ \textit{a priori arbitrary }parameters 
$a_{j},b_{jk}$ (or, equivalently, $\alpha _{j},\beta _{jk}$); this of course
implies that the $12$ coefficients $D_{k}$, $E_{jk}$ and $F_{j\ell }$ are
somewhat \textit{constrained}, presumably only $6$ of them might be \textit{%
arbitrarily} assigned.

While this indicates that the information on the behavior of the solutions
of the recursions (1)---for instance the fact that (almost) \textit{all} its
solutions starting from initial data in the interval such that $%
0<y_{j}\left( 0\right) <1$ are \textit{asymptotically periodic} with period $%
2$ (as described above)---shall also hold for the \textit{corresponding}
solutions of the system of $2$ \textit{nonlinear} recursions (7): 
\begin{equation}
\underset{n\rightarrow \infty }{Lim}\left[ x_{j}\left( n+2\right)
-x_{j}\left( n\right) \right] =0~;  \label{9}
\end{equation}%
a remarkable finding. Moreover it is plain from what we know (see above)
about the evolution of the system (1), that the system (7) features $4$
solutions periodic with period $2,$ $x_{j}\left( n+2\right) =x_{j}\left(
n\right) $, yielded by the $4$ \textit{initial} data%
\begin{eqnarray}
x_{j}\left( 0\right) &=&a_{j}~,  \notag \\
x_{j}\left( 0\right) &=&a_{j}+b_{j1}~,  \notag \\
x_{j}\left( 0\right) &=&a_{j}+b_{j2}~,  \notag \\
x_{j}\left( 0\right) &=&a_{j}+b_{j1}+b_{j2}~;  \label{10}
\end{eqnarray}%
and any interested reader may as well ascertain additional features of the
behavior of its solutions for \textit{other} assignments of the \textit{%
initial} data from the known behavior of the corresponding system (1) (see
[1]).

Let us finally emphasize that this system (7) is a more general \textit{%
discrete-time} version of the \textit{continuous-time} famous \textit{model}%
, which Vito Volterra invented (a century ago) to describe the interaction
of $2$ different kinds---prey and predators---of fishes in the sea, and
which is at the origin of many subsequent important developments in \textit{%
nonlinear }mathematics (see, for instance, the recent paper [4] and
references therein).

\bigskip

\textbf{Envoy}

Of course the "change of variables" approach outlined above is easily
extendable to the case of more than only $2$ dependent variables. It was
used by myself, alone or with collaborators, many times in the past, as well
as, over time, by very many other scientists worldwide; as it happens, now
the major production of interesting papers in this specific area seem to be
produced in China, mainly of course by Chinese colleagues, and as well by
non-Chinese colleagues working there; it has of course a long mathematical
tradition going back some centuries; while the work of the Italian
mathematician Vito Volterra mentioned above played a significant role in the
past century and continues to be relevant especially for applications of
this kind of mathematics to other scientific disciplines (ecology,
economics, epidemics,...); as well possibly for the actual realization of
certain \textit{mechanisms} meant to \textit{eventually} behave \textit{%
appropriately}, for instance\textit{\ periodically}, or at least \textit{not}
to yield possibly \textit{unwelcome}---\textit{diverging} or \textit{%
vanishing}---outcomes.

And finally---to justify the choice to focus on the mathematical
investigation of systems of \textit{nonlinear recurrences} rather than on
systems of \textit{nonlinear} \textit{ODEs} or \textit{PDEs}---the fact
should not be forgotten that \textit{all applicable} mathematical models
shall---most likely---hereafter eventually involve some use of \textit{%
electronic computers}, and that these instruments intrinsically use \textit{%
rational} numbers rather than \textit{real} numbers, and thereby eventually 
\textit{de facto} replace \textit{ODEs} or \textit{PDEs} with \textit{%
recurrences}:\textit{\ }so the study of \textit{recurrences }rather than 
\textit{ODEs} or\textit{\ PDEs} should maybe hereafter be the \textit{natural%
} tool to be utilized by whoever works to produce \textit{mathematical}
findings eventually \textit{applicable} to better understand \textit{nature}
as well as how our \textit{human world} functions, and thereby \textit{%
hopefully} to benefit \textit{humanity}.

$\bigskip $

\textbf{References}

[1] F. Calogero, "Simple recursions displaying interesting evolutions",

arXiv:2405.00370v1 [nlin.SI] 1 May 2024.

[2] F. Calogero, "Solvable nonlinear systems of 2 recursions displaying
interesting evolutions",

arXiv:2407.18270v1 [nlin.SI] 20 Jul 2024.

[3] F. Calogero, "Interesting system of 3 first-order recursions",

arXiv:2409.05074v1 [nlin.SI] 8 Sep 2024.

[4] M. Scalia, O. Ragnisco, B. Tirozzi, F. Zullo, "The Volterra integrable
case. Novel analytical and numerical results",

arXiv:2407.09155v2 [nlin.SI] 16 Jul 2024; Open Communications in Nonlinear
Mathematical Physics ]OCNMP[ Vol. 4 (2024) pp 188-212.

\end{document}